\documentclass[aps,prb,reprint,twocolumn,showpacs,groupeaddress,floatfix]{revtex4-2}
\usepackage{graphicx}  
\usepackage{dcolumn}   
\usepackage{bm}        
\usepackage{amssymb}   
\usepackage{amsmath}   

\usepackage{verbatim}  
\usepackage{xcolor}    

\usepackage{physics}
\usepackage{microtype}
\usepackage[colorlinks=true,allcolors=blue]{hyperref}
\usepackage[nameinlink,noabbrev]{cleveref}

\newcommand{\cS}{\mathcal{S}}
\newcommand{\cO}{\mathcal{O}}

\newcommand{\be}{\begin{equation}}
\newcommand{\ee}{\end{equation}}
\newcommand{\ba}{\begin{align}}
\newcommand{\ea}{\end{align}}
\newcommand{\sgn}{\operatorname{sgn}}

\begin{document}
\title{Suppression of quantized heat flow by the dielectric response of a compressible strip \\ at the quantum Hall edge}

\author{Eugene V. Sukhorukov}
\affiliation{Department of Physics, University of Geneva, CH-1211 Geneva, Switzerland}
\author{Adrien Tomà}
\affiliation{Department of Physics, University of Geneva, CH-1211 Geneva, Switzerland}

\date{\today}

\begin{abstract}
We develop a unified perturbative framework for energy transport along a chiral
quantum Hall (QH) edge coupled to a disordered, compressible strip. Treating
the strip as a generic linear-response environment characterized by its
retarded susceptibility $\chi_q^R(k,\omega)$, we obtain leading corrections to
both the heat flux carried by the edge plasmon and to its spectrum. Two generic
regimes emerge:  
(i) a gapped, local dielectric response with finite-range coupling, producing a
negative correction to the quantized heat flux that scales as $T^{4}$ at low
temperatures together with a convex cubic shift of the plasmon dispersion; and  
(ii) a hydrodynamic (diffusive) response with relaxation, yielding a crossover
from $T^{4}$ to $T^{3/2}$ scaling and a change of sign in the correction.
We further introduce a microscopic dipolar model in which the edge couples
electrostatically to localized dipole moments inside a wide compressible
strip.  This long-range interaction amplifies the nonlocal dielectric
back-action and generates new suppression laws, $T^{3}$, or even $T^{2}$ for
smooth disorder profiles, together with a universal ratio connecting
spectral curvature to thermal response.  
Across all regimes, the {\em total}  heat flux remains quantized: the apparent
deficit of the plasmon contribution reflects a reversible heat drag into the
compressible strip rather than a breakdown of quantization.   The
framework thus provides a coherent and quantitatively plausible explanation of
the “missing heat flux’’ anomaly and unifies the thermal and spectral
signatures of QH edge dynamics.
\end{abstract}


\maketitle

\section{Introduction}

Quantum Hall (QH) edge channels have long been regarded as perfectly ballistic and dissipationless conductors.
In the microscopic picture of drifting Landau level edge states~\cite{halperin_quantized_1982,Buttiker1988Nov} and in the effective chiral Luttinger-liquid description~\cite{Wen1990Jun,FROHLICH1991517}, they support unidirectional propagation of charge and energy without back-scattering or relaxation.
This view has made QH edges a paradigmatic realization of conservative one-dimensional transport.
Recent high precision experiments, however, have demonstrated clear departures from this ideal behavior.
Phase coherence measurements in Mach-Zehnder interferometers have shown progressive dephasing and partial loss of single-electron coherence even in the integer regime~\cite{neder_unexpected_2006,roulleau_direct_2008,litvin_decoherence_2007,bieri_finite-bias_2009,Roche1}, while more recent electron quantum optics experiments have revealed relaxation and decoherence of single-electron wave packets propagating along QH edges~\cite{Jullien,Fletcher2019,Bocquillon2013}.
Complementary charge- and energy-spectroscopy studies~\cite{Fujisawa2,Fujisawa3,Rodriguez2020,Rosenblatt} have further demonstrated incomplete thermalization and partial energy loss, even in nominally ballistic regimes.
Together, these findings indicate that QH edges, though chiral, are not isolated ballistic waveguides but interacting and dissipative one-dimensional systems.

A particularly active direction concerns heat transport.
Injecting energy into an edge channel through a biased quantum point
contact, quantum dot, or mesoscopic Ohmic contact generates
non-equilibrium states whose downstream evolution can be probed by
energy spectroscopy or noise
measurements~\cite{Altimiras2010,AltimirasRelaxation,Sueur,Sivre2018}.
These experiments have demonstrated incomplete equilibration between
modes, slow relaxation, and, most notably, a persistent deficit in the
measured energy current.
The resulting \emph{missing heat flux} anomaly, in which the thermal
conductance falls below its universal quantized
value~\cite{Sueur}, shows that the Tomonaga-Luttinger liquid
picture of an ideal chiral channel~\cite{Wen1990Jun,FROHLICH1991517,GiamarchiBook}
is insufficient to account for energy dissipation and relaxation at the
QH edge.

\begin{figure}[t]
  \centering
  \includegraphics[width=0.98\linewidth]{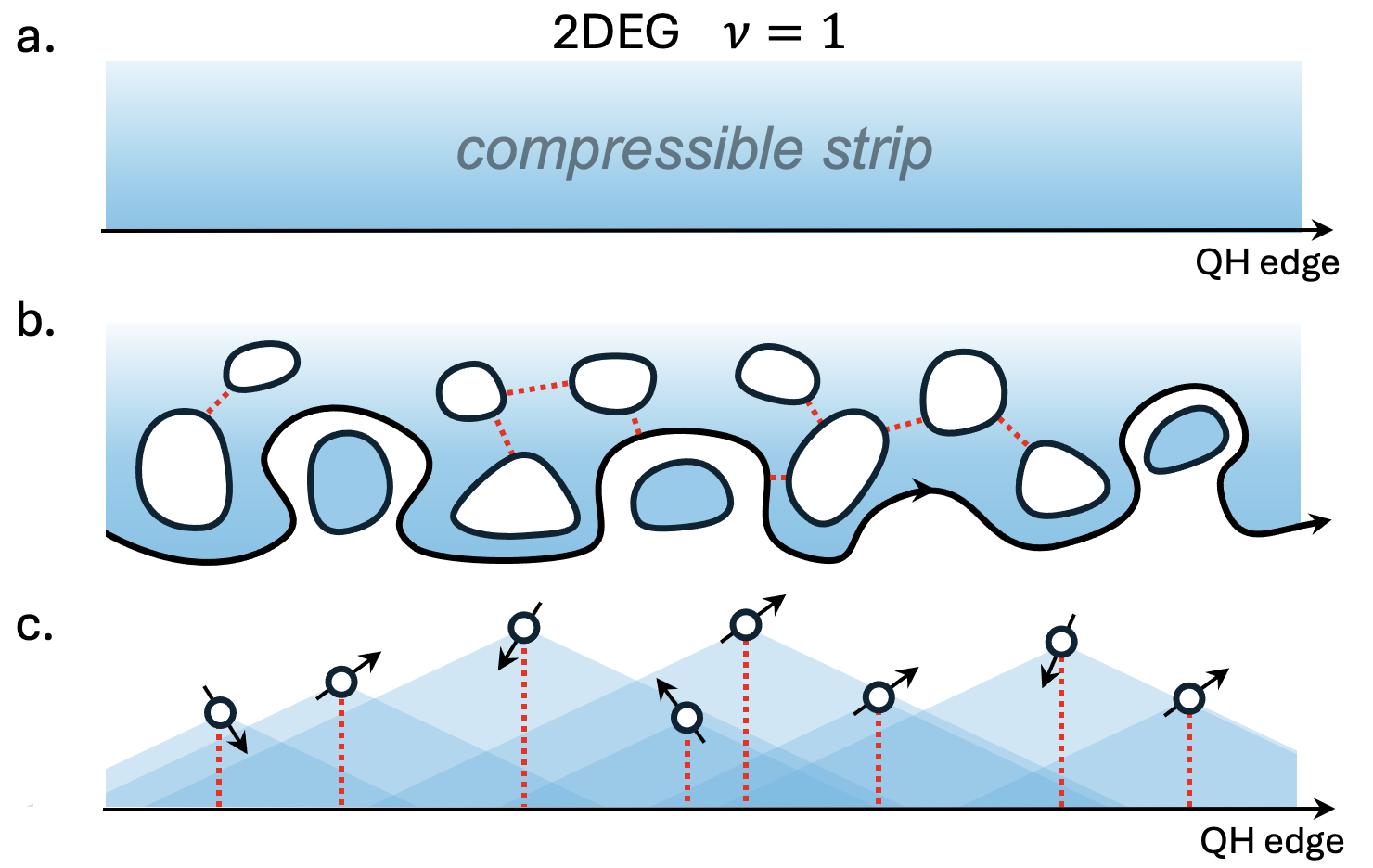}
  \caption{Models of dissipation and disorder at a quantum Hall (QH) edge. 
(a) Phenomenological description in which the compressible strip (gradient-shaded region) acts as a dissipative medium 
coupled to the chiral plasmon at the QH edge (solid line). 
(b) Microscopic picture of the compressible strip as a network of closed loops (“QH puddles”) 
connected by tunneling (dotted red lines) and capacitively coupled to the QH edge. 
(c) TLS model of the compressible strip, where two-level systems are randomly coupled to the edge (red dotted lines) 
or interact with the edge density field via a long-range potential (shaded triangles), 
resulting in a self-averaged dielectric response.}
  \label{fig:TLmodels}
\end{figure}

To address these puzzles, we recall that dissipation at the QH edge has
been attributed to the \emph{compressible strip} that forms near the
boundary of the two-dimensional electron gas
(Fig.~\ref{fig:TLmodels}a)~\cite{PhysRevB.46.4026}.  
At the hydrodynamic level~\cite{PhysRevLett.72.2935,goremykina_heat_2019},
fluctuations within this strip have been modeled as a
dissipative medium responsible for the relaxation of the edge plasmon.  
Such approaches capture irreversible energy flow but neglect the
disorder that inevitably characterizes the strip and introduces a
new microscopic length scale: the correlation length of potential
fluctuations.  
This scale can strongly affect both the coupling to the edge and the
nature of dissipation, motivating more microscopic descriptions in which
the strip is viewed as a disordered network of localized states or
``QH puddles'' capacitively coupled to the edge 
(Fig.~\ref{fig:TLmodels}b).

At elevated temperatures or short plasmon wavelengths, such networks act
effectively as continuous reservoirs, forming the basis for
transmission line (TL) models in which the compressible strip
is represented as a chain of metallic islands coupled to the edge
plasmon~\cite{Stabler,stabler_nonlocal}.  
While TL models provide a controlled route to Ohmic dissipation and
preserve thermodynamic consistency, their quantitative predictions for
thermal drag and heat flow~\cite{Sueur} disagree in sign with
experiment, suggesting that the dominant mechanism may instead originate
from the intrinsic, frequency-dependent dielectric response of the
compressible strip itself.

At lower energies, where only a few localized states are active, the
microscopic picture naturally reduces to a sparse set of low-energy
fluctuators coupled to the edge.  
Such fluctuators can be modeled as quantum two-level systems (TLSs),
providing a minimal microscopic realization of the strip’s
polarization response (Fig.~\ref{fig:TLmodels}c).  
TLS-type fluctuators are a well established paradigm for dissipation in
disordered solids~\cite{Anderson01011972,Phillips1972} and have been
directly implicated in the decoherence of superconducting qubits and
mesoscopic devices~\cite{PhysRevLett.95.210503,PhysRevLett.93.077003,PhysRevLett.93.180401}.
In the QH regime, scanning tunneling microscopy has revealed localized
electronic states and Landau level tails within the
compressible region~\cite{Morgenstern2003,Hashimoto2008}, while more
recent experiments in graphene identified trap states consistent with
two-level fluctuators that couple to edge
transport~\cite{CHAU202126}.  
Hence, the TLS model offers a convenient microscopic realization of a generic dielectric environment without assuming a
specific origin of disorder.

In this paper, however, we do not immediately rely on a specific microscopic model
of disorder in the compressible strip.  
Instead, we formulate a general perturbative framework describing the
coupling between a chiral QH edge plasmon and a disordered,
compressible environment characterized by an arbitrary linear response (dielectric or hydrodynamic).  
Within this approach, the leading corrections to both the
edge heat flux and the plasmon spectrum are expressed entirely through
the retarded susceptibility $\chi_q^R(k,\omega)$ of the strip.  
Two complementary realizations are analyzed in detail:  
(i) a gapped, local dielectric response with finite-range coupling,
which produces a universal negative $T^4$ correction to the quantized
heat flux and a corresponding convex cubic correction to the plasmon
dispersion; and  
(ii) a hydrodynamic (diffusive) response with optional relaxation,
leading to richer scaling laws, from $T^4$ to $T^{3/2}$, together with
a crossover in the sign of the correction.  
In both cases, the environmental back action, dielectric or hydrodynamic, reduces the group velocity of the edge plasmon and can suppress the thermal conductance by an amount
comparable to that observed experimentally~\cite{Sueur}.  

Beyond these two generic response types, we also develop in
Sec.~\ref{Microscopic model} a microscopic dipolar model in which the edge
plasmon couples electrostatically to localized dipole moments inside the
compressible strip.  
This long-range interaction enhances the nonlocal dielectric back action and
leads to additional regimes of heat-flux suppression.  
For a narrow strip, the dipolar model reproduces the $T^{4}$
correction obtained in the finite-range TLS scenario, whereas for a wide strip
the transverse integration over many dipolar layers generates a 
$T^{3}$ law; if the disorder density varies smoothly across the strip, the
power law can soften further to $T^{2}$.  
Moreover, the dipolar scenario yields its own characteristic relation between
spectral curvature and heat-flux suppression, complementing the universal
relations found in the finite-range TLS and diffusive regimes.  
These additional signatures provide a microscopic route for identifying the
dominant mechanism behind the missing-heat anomaly in experiment.

It is important to emphasize that the quantization of the \emph{total}
heat flux at the QH edge is not violated.  
The correction $\Delta J_{\rm th}$ refers only to the energy current
carried by the \emph{edge plasmon}, whose coupling to the compressible
strip redistributes heat between the chiral channel and the surrounding
disordered environment.  
The ``missing'' portion of the plasmon heat flux thus reappears as an
induced energy flow within the strip itself, representing a form of
\emph{heat drag} between the two subsystems.  
Depending on the sign of the dielectric back action, this drag can be
positive or negative.  
Hence, the observed suppression of the plasmon contribution does not
signify a breakdown of quantization, but rather a reversible exchange of
energy between coupled chiral and nonchiral modes.

Although our analysis focuses on a single chiral channel, it directly applies to the experimentally relevant case of filling factors $\nu>1$, particularly $\nu = 2$, where two co-propagating edge modes exist.
In that regime, long-range Coulomb interactions diagonalize the dynamics into a fast charge mode and a slow dipole (neutral) mode~\cite{LevkivskyiSukhorukov2008,Degiovanni2009}.
The charge mode, being symmetric in the two edges, couples weakly to disorder and remains effectively ballistic, preserving its quantized contribution to the thermal conductance.
In contrast, the dipole mode is more strongly coupled to the surrounding compressible strip, which can polarize in response to its local electric field.
It is therefore natural to associate the single plasmon mode in our model with this dipole branch.

\section{Model of QH edge}
\label{sec:model}

We consider a single chiral quantum Hall (QH) edge  mode (right-moving plasmon) interacting with a disordered, compressible medium localized near the edge (``environment'') representing a compressible strip. The environment will be kept \emph{generic} in this section: we do not assume any specific microscopic Hamiltonian beyond linear response. Our goal is to (i) define a minimal Hamiltonian description with a finite–range edge to environment coupling compatible, and (ii) fix the expression for the energy current (heat flux) carried by the edge in the presence of the coupling. Microscopic forms (independent TLSs, etc.) will be specified later as particular realizations of the same linear response framework. Throughout this work we set $e=\hbar=1$, so that all velocities, energies, and
couplings are expressed in natural units.

We split the Hamiltonian 
\begin{equation}
H=H_0+H_{\rm int},\qquad H_0\equiv H_{\rm edge}+H_{\rm env},
\label{eq:ham}
\end{equation}
and work in the interaction picture generated by $H_0$. All operators carry the $H_0$ induced time evolution, while the interaction $H_{\rm int}$ is treated perturbatively. The environment is a disordered, compressible strip with Hamiltonian $H_{\rm env}$. We only require that its operators admit well-defined equilibrium retarded correlators so that linear response applies. No further structure is needed at this stage.

The edge is described by a chiral boson field $\phi(x,t)$ with  velocity $v>0$ and Hamiltonian
\begin{subequations}
\label{eq:H-comm}
\begin{align}
&H_{\rm edge}=\frac{v}{4\pi}\int dx\,\big(\partial_x\phi(x)\big)^2,
\label{eq:edge-H}\\
&[\partial_x\phi(x),\phi(y)]=2\pi i\,\delta(x-y).
\label{eq:edge-comm}
\end{align}
\end{subequations}
The equation of motion follows from the commutation relation  
\eqref{eq:edge-comm}$, (\partial_t+v\partial_x)\phi=0$, so excitations propagate towards  $x=+\infty$.

In the linear response framework, the most general coupling is to the local edge charge density $\rho(x,t)=(1/2\pi)\partial_x\phi(x,t)$:
\begin{equation}
H_{\rm int}(t)=\frac{1}{2\pi}\int dx\;\partial_x\phi(x,t)\;q(x,t),
\label{eq:Hint-fr}
\end{equation}
where the density field $q(x,t)$ collects the microscopic couplings to local environment operators $S_i(t)$ (e.g., $S_i=\sigma_i^z$ for TLSs) placed at random positions $x_i$:
\begin{equation}
q(x,t)\equiv \sum_i g_i(x-x_i)\,S_i(t).
\label{eq:q-def}
\end{equation}
The profiles $g_i(x-x_i)$ encode both the spatial range and the randomness of the edge-environment interaction. 
As illustrated in Fig.~\ref{fig:TLmodels}(c), this construction corresponds to a disordered array of localized fluctuators (e.g., TLSs) coupled to the edge field either directly or through finite-range interactions, whose collective response is described by the coarse-grained field $q(x,t)$. 
Two limits of \eqref{eq:q-def} will be used repeatedly:

\noindent (a) Discrete (point-like) coupling: $g_i(x-x_i)=g_i\,\delta(x-x_i)$, which recovers
\begin{equation}
H_{\rm int}(t)=\sum_i \frac{g_i}{2\pi}\,\partial_x\phi(x_i,t)\,S_i(t).
\label{eq:Hint-point}
\end{equation}
\noindent (b) Finite-range, homogeneous coupling: $g_i(x-x_i)=g\,f(x-x_i)$ with a real, even kernel $f$ of range $x_0$ and $\int dx\,f(x)=1$. Then $q(x,t)=g\sum_i f(x-x_i)S_i(t)$; disorder averages will enter through overlap kernels built from $f$.

The energy density of the edge is $\mathcal{H}_{\rm edge}(x)=(v/4\pi)\,[\partial_x\phi(x)]^2$. Using the Heisenberg equation with $H_{\rm edge}$ and the equal-time commutator in \eqref{eq:edge-comm}, one obtains the continuity equation $\partial_t \mathcal{H}_{\rm edge}+\partial_x j_E=0$ with the energy current density
\begin{equation}
j_E(x,t)=\frac{v^2}{4\pi}\,\big[\partial_x\phi(x,t)\big]^2.
\label{eq:jE}
\end{equation}
In a stationary state, we evaluate the heat flux at a fixed point (e.g., $x=0$): $J_{\rm th}\equiv\langle j_E(0)\rangle$. Expressed via the Keldysh correlator $G^K(0,\omega)=\int dt\,e^{i\omega t}\langle\{\phi(0,t),\phi(0,0)\}\rangle$, one finds
\begin{equation}
J_{\rm th}
=\frac{1}{8\pi}\int_{-\infty}^{\infty}\frac{d\omega}{2\pi}\;\omega^2\,G^{K}(0,\omega),
\label{eq:J-from-GK}
\end{equation}
where the vacuum ($T=0$) piece must be subtracted. For the \emph{free} chiral boson in equilibrium at temperature $T$ [with $G_0^K(0,\omega)=\frac{2\pi}{\omega}\coth(\beta\omega/2)$, see Appendix \ref{app:FDT}], Eq.~\eqref{eq:J-from-GK} gives the heat flux  equal to 
\begin{equation}
J_q=\frac{\pi}{12}\,T^2,
\label{eq:Jq}
\end{equation}
i.e., the heat flux quantum.
All interaction effects thus enter as a correction $\Delta J_{\rm th}\equiv J_{\rm th}-J_q$, which we will compute to $\mathcal{O}(g^2)$ in the next section.

Because the edge is chiral, only environment operators located \emph{upstream} can influence the edge field at a given point. In practice this appears as an $x<0<x'$ selection inside the spatial integrals that define the $\mathcal{O}(g^2)$ back action. The upstream (same-side) sector cancels in equilibrium by the fluctuation-dissipation theorem FDT, ensuring that the theory respects causality and detailed balance. We will make this selection explicit in Sec.~\ref{sec:deltaJ-formal}.

\section{Formal expression for the heat flux correction}
\label{sec:deltaJ-formal}

We now derive the $\mathcal O(g^2)$ correction to the edge heat flux within the interaction picture set by $H_0$ and the coupling in Eq.~\eqref{eq:Hint-fr} and \eqref{eq:q-def}. The key object is the Keldysh correlator at the probe point $x=0$,
\begin{equation}
G^{K}(0,\omega)=\!\int\!dt\,e^{i\omega t}\,\big\langle\{\phi_H(0,t),\phi_H(0,0)\}\big\rangle,
\end{equation}
which enters the heat current via Eq.~\eqref{eq:J-from-GK}. We assume a factorized initial state at $t_0=-\infty$, $\rho_0=\rho_{\rm edge}\otimes\rho_{\rm env}$, with any static shift subtracted so that $\langle q(x)\rangle=0$ (equivalently, $\langle S_i\rangle=0$). We expand the Heisenberg field $\phi_H$ to $\mathcal O(g^2)$ 
\begin{align}
\phi_H(x,t)=\phi(x,t)+i\!\int_{-\infty}^t\!dt_1\,[H_{\rm int}(t_1),\phi(x,t)]
\nonumber \\
-\!\int_{-\infty}^t\!dt_1\!\int_{-\infty}^{t_1}\!dt_2\,[H_{\rm int}(t_2),[H_{\rm int}(t_1),\phi(x,t)]],
\label{app:eq:phi-Dyson}
\end{align}
and use the commutator \eqref{eq:commutator}.
Collecting terms yields the compact commutator with all nontrivial operator content residing in environment
correlators 
\begin{subequations}
\label{eq:chiq-def}
\begin{align}
&\chi_q^{R/A}(x,x';t)\equiv \mp\,i\,\theta(\pm t)\,\big\langle[\,q(x,t),q(x',0)\,]\big\rangle,
\label{eq:chiq-def-RA}\\[0.5em]
&\chi_q^K(x,x';t)\equiv \big\langle\{\,\delta q(x,t),\delta q(x',0)\,\}\big\rangle.
\label{eq:chiq-def-K}
\end{align}
\end{subequations}
and the edge appearing only via the $c$-number correlator $G_0^R$.
.

To second order, $G^K$ receives (i) an upstream-upstream piece (both interaction vertices on the same, left side of the probe) and (ii) a genuine back action piece that links the left ($x<0$) and right ($x'>0$) sides. In frequency space this can be written compactly 
\begin{widetext}
\begin{align}
\delta^{(2)}G^{K}(0,\omega)
&=\underbrace{\frac{1}{v^{2}}\!\int_{-\infty}^{0}\!dx\!\int_{-\infty}^{0}\!dx'\;
e^{-i\frac{\omega}{v}(x-x')}\!
\left[\chi_q^{K}(x,x';\omega)-i\,\coth\!\Big(\frac{\beta_e\omega}{2}\Big)\big(\chi_q^{R}-\chi_q^{A}\big)\!(x,x';\omega)\right]}_{\text{upstream--upstream}}
\nonumber\\[0.4em]
&\quad+\underbrace{\frac{2}{v^{2}}\,\coth\!\Big(\frac{\beta_e\omega}{2}\Big)\!
\int_{-\infty}^{0}\!dx\!\int_{0}^{\infty}\!dx'\;
\Big[\cos\!\Big(\tfrac{\omega}{v}(x-x')\Big)\,\mathrm{Im}\chi_q^{R}
-\sin\!\Big(\tfrac{\omega}{v}(x-x')\Big)\,\mathrm{Re}\chi_q^{R}\Big]\!(x,x';\omega)}_{\text{back–action}}.
\label{eq:deltaG-structure}
\end{align}
\end{widetext}
If the environment and the edge share the same temperature $T$ (equilibrium), the FDT,
$\chi_q^{K}=i\,\coth(\beta\omega/2)\,(\chi_q^{R}-\chi_q^{A})$, cancels the upstream-upstream block in Eq.~\eqref{eq:deltaG-structure}. The entire $\mathcal O(g^2)$ effect then comes from the \emph{upstream$\to$downstream} back action and can be presented in a manifestly causal form,
\begin{align}
&\delta^{(2)}G^{K}(0,\omega)
=\frac{2}{v^{2}}\,\coth\!\Big(\frac{\beta\omega}{2}\Big)\nonumber\\
&\quad\times\int_{-\infty}^{0}\!dx\!\int_{0}^{\infty}\!dx'\;
\mathrm{Im}\!\left\{e^{-i\frac{\omega}{v}(x-x')}\,\chi_q^{R}(x,x';\omega)\right\}.
\label{eq:deltaG-backaction}
\end{align}

Inserting this correction into Eq.~\eqref{eq:J-from-GK} and subtracting the $T=0$ part gives the compact, universal form
\begin{equation}
\Delta J_{\rm th}
=\int_{-\infty}^{\infty}\frac{d\omega}{8\pi^2}\;\omega^2\,
\Big[\coth\!\Big(\frac{\beta\omega}{2}\Big)-\mathrm{sgn}\,\omega\Big]\,
\mathcal S(\omega),
\label{eq:DJ-master}
\end{equation}
where we define the source kernel
\begin{equation}
\mathcal S(\omega)\equiv\frac{1}{v^2}\!\int_{-\infty}^{0}\!\!dx\!\int_{0}^{\infty}\!\!dx'
\mathrm{Im}\!\left\{e^{-i\frac{\omega}{v}(x-x')}\chi_q^{R}(x,x';\omega)\right\}.
\label{eq:S-kernel}
\end{equation}
This general results is the starting point for all specializations below.

By linearity of $q$ in Eq.~\eqref{eq:q-def}, the retarded function factorizes into coupling profiles and environment susceptibilities,
\begin{equation}
\chi_q^R(x,x';\omega)
=\sum_{i,j} g_i(x-x_i)\,g_j(x'-x_j)\;\chi^R_{ij}(\omega),
\label{eq:chiq-build}
\end{equation}
where $\chi^R_{ij}(\omega)$ are the site-resolved environment (e.g., TLS) retarded correlators
\begin{equation}
\chi^R_{ij}(\omega)=-i\int_0^\infty dt\, e^{i\omega t}\,\big\langle[\,S_i(t),S_j(0)\,]\big\rangle.
\label{eq:chi-R}
\end{equation}
Equation~\eqref{eq:S-kernel} together with \eqref{eq:chiq-build} yields a fully continuous expression for $\mathcal S(\omega)$ that automatically reduces to the homogeneous, discrete, and finite-range limits.

\medskip
\noindent\emph{(a) Homogeneous
medium},  $\chi_q^R(x,x';\omega)=\chi_q^R(x{-}x',\omega)$ and the upstream/downstream
selection reduces to a single separation variable
$r=x'-x>0$, giving
\begin{equation}
\cS(\omega)
=\frac{1}{v^2}\int_{0}^{\infty}\!r\,dr\;
\mathrm{Im}\!\left\{e^{-i(\omega/v)r}\,\chi_q^{R}(-r,\omega)\right\}.
\label{eq:S-realspace}
\end{equation}

\medskip
\noindent\emph{(b) Pointlike (discrete) coupling} with $g_i(x-x_i)=g_i\,\delta(x-x_i)$,
\begin{equation}
\mathcal S(\omega)
=\frac{1}{v^2}\sum_{i,j} g_i g_j\;\theta(-x_i)\theta(x_j)\;
\mathrm{Im}\!\left\{e^{-i\frac{\omega}{v}(x_i-x_j)}\,\chi^R_{ij}(\omega)\right\}.
\label{eq:S-discrete}
\end{equation}

\medskip
\noindent\emph{(c) Finite-range, homogeneous coupling to independent sites}, and assuming site-diagonal response (Fig.~\ref{fig:TLmodels}c)
\begin{equation}
g_i(x-x_i)=g\,f(x-x_i),\qquad
\chi^R_{ij}(\omega)=\delta_{ij}\,\chi^R_{\rm loc}(\omega),
\label{eq:local}
\end{equation}
where $f$ real, even function, and normalized $\int dx\,f=1$. Assuming the range of interaction $x_0$ to be relatively long, disorder averaging over positions with density $n_s$ gives
\begin{subequations}
\label{eq:chi-loc}
\begin{align}
\label{eq:chi-loc-C}
\chi^R_q(x-x',\omega)&=g^2\,n_s\,\chi_{\rm loc}^R(\omega)C(x-x'),\\[0.3em]
C(r)\!&\equiv\!\int dxf(x)f(x+r),
\label{eq:C-FC}
\end{align}
\end{subequations}
where $C(r)$ is the even, positive overlap function. Substituting these expressions to Eq.~\eqref{eq:S-kernel}, we obtain for $n_sx_0\gg 1$
\begin{subequations}
\label{eq:S}
\begin{align}
\mathcal{S}(\omega)&=\frac{n_s g^2}{v^2}\,
\mathrm{Im}\!\left\{\chi_{\rm loc}^R(\omega)\,{\cal F}(\omega)\right\},
\label{eq:S-finiterange}
\\[0.4em]
{\cal F}(\omega)&=\int_{0}^{\infty}\!\!r\,drC(r)e^{i(\omega/v)r}.
\label{eq:S-F}
\end{align}
\end{subequations}

\medskip
These equations constitute the general $\mathcal O(g^2)$ result: the entire problem is reduced to the retarded correlator $\chi_q^R$ of the environment (compressible strip), evaluated between points upstream and downstream of the probe and modulated by the chiral phase factor $e^{-i\omega(x-x')/v}$. All model dependence enters solely through $\chi_q^R$ (or, equivalently, through $\chi^R_{ij}$ and the coupling profiles $g_i$). This form will be the basis for the low-$T$ analysis in Secs.~\ref{sec:diffusion} and \ref{sec:tls-finite-range}.

\section{Plasmon self-energy and spectrum}
\label{sec:spectrum}

In this section we derive the dressed retarded propagator of the edge plasmon and extract the
on-shell dispersion and damping, working entirely with the density field
$q(x,t)$ defined in Eq.~\eqref{eq:q-def} and the interaction Hamiltonian \eqref{eq:Hint-fr}. All environment details enter only through the retarded correlator of $q$ defined in Eq.~\eqref{eq:chiq-def-RA}.
Starting from the second order expansion of the Heisenberg field \eqref{app:eq:phi-Dyson} and proceeding as in Sec.\ \ref{sec:deltaJ-formal},
one can express the $\cO(q^2)$ correction to the retarded propagator
\begin{equation}
G^{R}(x,t)=-i\theta(t)\langle[\phi_H(x,t),\phi_H(0,0)]\rangle
\end{equation}
as a convolution of free edge correlators with the retarded correlator of the environment
A short rearrangement of the three $\cO(q^2)$ blocks (cf.\ the heat flux calculation) yields
\begin{align}
\delta G^{R}(x,t)
&=\frac{1}{(2\pi)^2}\!\int\!dx'dx''\!\int\!dt_1dt_2\;
G_0^R(x-x',t-t_1)\nonumber\\[0.3em]
&\times \partial_{x'}\partial_{x''}\, \chi_q^R(x',x'';t_1-t_2)\;G_0^R(x'',t_2).
\label{eq:s3-Dyson-xt}
\end{align}
Using translation invariance after the disorder averaging,
namely, replacing $\chi_q^{R}(x,x';t)\to \chi_q^{R}(x-x';t)$, one finds
\begin{align}
G^{R}(k,\omega)
\!&=\!G_0^{R}(k,\omega)\nonumber \\[0.3em]
&+(2\pi)^{-2}k^2G_0^{R}(k,\omega)\chi_q^{R}(k,\omega)G_0^{R}(k,\omega).
\label{eq:s3-Dyson}
\end{align}

Resumming the geometric series, the dressed retarded Green function reads
\begin{equation}
G^{R}(k,\omega)=\frac{G_0^{R}(k,\omega)}{1-(2\pi)^{-2}\,\chi_q^{R}(k,\omega)G_0^{R}(k,\omega)}. 
\label{eq:s3-GR-dressed}
\end{equation}
Using the free chiral propagator \eqref{app:eq:GRGA-omega-a} (Appendix~\ref{app:FDT}) in Fourier space
\begin{equation}
G_0^{R}(k,\omega)=\frac{2\pi}{k}\,\frac{1}{\omega-vk+i0^+},
\label{eq:s3-GR0}
\end{equation}
we can equivalently write
\begin{equation}
G^{R}(k,\omega)=\frac{2\pi}{k}\,\frac{1}{\omega-vk-\Sigma^{R}(k,\omega)}
\label{eq:s3-GR}
\end{equation}
with the self-energy
\begin{equation}
\Sigma^{R}(k,\omega)=\frac{k}{2\pi}\chi_q^{R}(k,\omega).
\label{eq:s3-GR-sigma}
\end{equation}
The plasmon pole is located at $\omega_k=vk+\delta\omega_k-i\Gamma_k$, with
\begin{subequations}
\label{eq:s3-onshell}
\begin{align}
\delta\omega_k&=\frac{k}{2\pi}\,\mathrm{Re}\,\chi_q^{R}(k,\omega)\big|_{\omega=vk},
\label{eq:s3-onshell-omega1}
\\[0.2em]
\Gamma_k&=-\,\frac{k}{2\pi}\,\mathrm{Im}\,\chi_q^{R}(k,\omega)\big|_{\omega=vk}.
\label{eq:s3-onshell-gamma1}
\end{align}
\end{subequations}

Equations~\eqref{eq:s3-onshell} are completely general: to obtain the spectrum and attenuation to $\cO(q^2)$, one needs only the retarded susceptibility $\chi_q^R(k,\omega)$ of the environment density field. In particular, using \eqref{eq:chiq-build} and \eqref{eq:local}
for finite-range, homogeneous coupling to independent sites, one obtains
\begin{subequations}
\label{eq:s3-onshell-local}
\begin{align}
\delta\omega_k&=\frac{1}{2\pi}g^2n_sk|\,\tilde f(k)\,|^2\,\mathrm{Re}\,
\chi_{\rm loc}^{R}(\omega)\big|_{\omega=vk},
\label{eq:s3-onshell-omega2}
\\[0.4em]
\Gamma_k&=-\frac{1}{2\pi}g^2n_sk|\,\tilde f(k)\,|^2\,\mathrm{Im}\,
\chi_{\rm loc}^{R}(\omega)\big|_{\omega=vk}.
\label{eq:s3-onshell-gamma2}
\end{align}
\end{subequations}
where $\tilde f(k)=\int dx\,e^{-ikx}f(x)$.

\section{Diffusive and relaxing medium}
\label{sec:diffusion}

In this section we specialize the general results of
Secs.~\ref{sec:deltaJ-formal}–\ref{sec:spectrum} to a homogeneous
``compressible strip'' whose long–wavelength dynamics is diffusive,
optionally regularized by a weak local relaxation channel. We work
throughout with the coupling density field $q(x,t)$ of
Eq.~\eqref{eq:q-def} and its retarded susceptibility $\chi_q^R$
introduced in Eq.~\eqref{eq:chiq-def-RA}; no additional microscopic
fields are introduced.

\subsection{Thermodynamic derivation of the hydrodynamic response}

Near local equilibrium, the coarse-grained free energy reads
\begin{equation}
F[q]=\int dx\,\bigg[\frac{q^2(x)}{2\,\chi_q^{0}}+q(x)\,\rho(x)\bigg],
\end{equation}
so that the thermodynamic force (``chemical potential'')
\begin{equation}
\mu_q(x,t)\equiv\frac{\delta F}{\delta q(x,t)}=\frac{q(x,t)}{\chi_q^{0}}+\rho(x,t),
\label{eq:mu-q}
\end{equation}
vanishes in equilibrium: $\mu_q=0$ gives $q=-\chi_q^0\rho$. Here, $\chi_q^{0}>0$ so that a positive $\chi_q^0$ corresponds to a normal, positive compressibility.

Linear irreversible thermodynamics closes the dynamics with the
linear response  relation and the continuity equation \cite{LandauKinetics,deGrootMazur}
\begin{subequations}
\label{eq:thermodynamics}
\begin{align}
&j_q(x,t)=-\,\sigma\,\partial_x\mu_q(x,t),\qquad
\sigma=\chi_q^{0}D,
\label{eq:constit-q}
\\[0.4em]
&\partial_t q(x,t)+\partial_x j_q(x,t)=-\,\gamma_s\,\chi_q^{0}\,\mu_q(x,t),
\label{eq:cont-q}
\end{align}
\end{subequations}
where the second equation  in \eqref{eq:constit-q} is the Einstein relation,  $D$ is the diffusion constant and $\gamma_s\!\ge\!0$ is an
optional weak (``slow'') local relaxation rate that regularizes the infrared
(set $\gamma_s=0$ for a strictly conserved $q$). 

Combining
Eqs.~\eqref{eq:mu-q}–\eqref{eq:cont-q} gives the driven diffusion equation
\begin{align}
\partial_t q(x,t)-D\,\partial_x^2 q(x,t)&+\gamma_s\,q(x,t)\nonumber
\\[0.3em]
&=-\chi_q^{0}\,\big(\gamma_s-D\,\partial_x^2\big)\,\rho(x,t).
\label{eq:diffusion-source-q}
\end{align}
Applying the Fourier transform one obtains,
\begin{equation}
\big(-i\omega+Dk^2+\gamma_s\big)\,q(k,\omega)
=-\chi_q^{0}\,(\gamma_s+Dk^2)\,\rho(k,\omega),
\end{equation}
and comparing to the Kubo form
$q=\chi_q^R\,\rho$ yields the retarded susceptibility
\begin{equation}
\chi_q^R(k,\omega)
=-\,\chi_q^{0}\,\frac{\gamma_s+Dk^2}{-\,i\omega+\gamma_s+Dk^2},
\label{eq:chiq-hydro}
\end{equation}
which satisfies $\chi_q^R(k,0)=-\chi_q^{0}$.   The real-space representation at fixed frequency is
\begin{subequations}
\label{eq:chiq-x}
\begin{align}
\chi_q^R(x,\omega)&=-\chi_q^{0}\,\delta(x)
-\frac{i\omega\,\chi_q^{0}}{2D\,\alpha_\omega}\,e^{-|x|\,\alpha_\omega},
\label{eq:chiq-x-x}
\\[0.5em]
\alpha_\omega&=\sqrt{(\gamma_s-i\omega+0^+)/D}.
\label{eq:chiq-x-omega}
\end{align}
\end{subequations}
For the following, it is convenient to introduce the relaxation length parameter $l_s=1/\alpha_0$.

\subsection{Heat flux kernel ${\cal S}(\omega)$ and $\Delta J_{\rm th}(T)$}

The general back action formula of Sec.~\ref{sec:deltaJ-formal} gives
the $\mathcal{O}(q^2)$ correction to $G^K(0,\omega)$ entirely in terms
of $\chi_q^R$.  Accordingly, the heat flux correction follows from
Eqs.~\eqref{eq:DJ-master} and \eqref{eq:S-realspace}. 
Using Eq.~\eqref{eq:chiq-x} (the $\delta$-term vanishes because of
the prefactor $r$), the integral is elementary and one obtains the
closed form
\begin{equation}
\cS(\omega)
=-\,\frac{\chi_q^{0}}{v^2}\;\frac{\omega}{2D}\;
\mathrm{Re}\!\left\{\frac{1}{\alpha_\omega\,\big(\alpha_\omega+i\omega/v\big)^2}\right\}.
\label{eq:S-closed}
\end{equation}
As a function of frequency, $\cS(-\omega)=-\cS(\omega)$.

Substituting \eqref{eq:S-closed} into the formula
\eqref{eq:DJ-master} and evaluating the integral asymptotically in $\omega$,
one finds three temperature windows governed by the scales
$\gamma_s$ and $\omega_*\equiv v^2/D$:
\begin{subequations}
\begin{align}
&\Delta J_{\rm th}(T)
=\;-\;\frac{\pi^{2}}{60}\;\frac{\chi_q^{0}\sqrt{D}}{\gamma_s^{3/2}v^2}\,
\;T^{4},
\qquad (\gamma_s,\omega_*\gg T),
\label{eq:DJ-lowT-q}
\\[0.4em]
&\Delta J_{\rm th}(T)
=\frac{3\,\zeta\!\big(\tfrac{5}{2}\big)}{16\sqrt{2}\,\pi^{3/2}}
\frac{\chi_q^{0}\sqrt{D}}{v^2}\,T^{5/2},
\; ( \omega_*\gg T\gg \gamma_s),
\label{eq:DJ-midT-q}
\\[-0.4em]
&\Delta J_{\rm th}(T)
=\frac{\zeta\!\big(\tfrac{3}{2}\big)}{8\sqrt{2}\,\pi^{3/2}}\;
\frac{\chi_q^{0}}{\sqrt{D}}\;T^{3/2},
\; (T\gg \gamma_s,\omega_*),
\label{eq:DJ-highT-q}
\end{align}
\end{subequations}
where $\zeta$ is the Reimann zeta function. At the very lowest temperatures the correction is \emph{negative} and
quartic in $T$; above the relaxation scale it changes sign and crosses
over to the diffusive power laws $T^{5/2}$ and $T^{3/2}$.

\subsection{Plasmon self-energy and spectrum in the diffusive medium}

The dressed retarded edge propagator and its self-energy were obtained
in Sec.~\ref{sec:spectrum}.
Inserting the hydrodynamic form \eqref{eq:chiq-hydro} into Eqs.~\eqref{eq:s3-onshell}  and
evaluating on shell $\omega=vk$ gives compact expressions for the
dispersion shift and linewidth:
\begin{subequations}
\begin{align}
\delta\omega_k
&= -\,\frac{\chi_q^{0}}{2\pi}\,k\,
\frac{(\gamma_s+Dk^2)^2}{(\gamma_s+Dk^2)^2+(vk)^2},
\label{eq:s4-dw}
\\[0.4em]
\Gamma_k
&= \frac{\chi_q^{0}}{2\pi}\,k\,
\frac{(\gamma_s+Dk^2)\,(vk)}{(\gamma_s+Dk^2)^2+(vk)^2}.
\label{eq:s4-gam}
\end{align}
\end{subequations}
Introducing $k_*\equiv v/D$ (equivalent to $\omega_*=v^2/D$ on shell), three limits follow:

\medskip
\noindent\emph{(a) Relaxation–regularized infrared:} $\gamma_s,\omega_*\gg \omega$, or equivalently, $\gamma_s/v,k_*\gg k$. 
The non-conserving channel cuts off the hydrodynamic singularity:
\begin{equation}
\delta\omega_k \simeq -\,\frac{\chi_q^{0}}{2\pi}\,
\left(k-\frac{v^2}{\gamma_s^2}\,k^3\right),
\qquad
\Gamma_k \simeq \frac{\chi_q^{0}}{2\pi}\,\frac{v}{\gamma_s}\,k^2.
\label{eq:s4-regC}
\end{equation}
Again the group velocity is reduced, and the spectrum is {\em convex}, while the damping is suppressed by
$vk/\gamma_s$ at small~$k$.

\medskip
\noindent\emph{(b) Drift-dominated on shell:} $ \omega_*\gg \omega\gg \gamma_s$. Here 
\begin{equation}
\delta\omega_k \simeq -\,\frac{\chi_q^{0}}{2\pi}\,\frac{D^2}{v^2}\,k^{3},
\qquad
\Gamma_k \simeq \frac{\chi_q^{0}}{2\pi}\,\frac{D}{v}\,k^{2}.
\label{eq:s4-regA}
\end{equation}
The leading dispersive correction is cubic ({\em concave} spectrum) and the
damping is quadratic in~$k$.

\medskip
\noindent\emph{(c) Diffusion-dominated on shell}: $ \omega\gg \gamma_s,\omega_*$. We obtain
\begin{equation}
\delta\omega_k \simeq -\,\frac{\chi_q^{0}}{2\pi}\,k\,[1-(v/Dk)^2],
\qquad
\Gamma_k \simeq \frac{\chi_q^{0}}{2\pi}\,\frac{v}{D}.
\label{eq:s4-regB}
\end{equation}
The group velocity is reduced and the spectrum is {\em convex}.  The linewidth
saturates.

\medskip
Equations~\eqref{eq:DJ-lowT-q}–\eqref{eq:DJ-highT-q} and
\eqref{eq:s4-dw}–\eqref{eq:s4-regC} provide the complete description of the diffusive regime and relaxation for both the heat–flux
correction and the plasmon spectrum within the $q$-field formulation.
All microscopic models that coarse-grain to
Eq.~\eqref{eq:chiq-hydro} must collapse to these universal forms in
their respective hydrodynamic windows.

\section{Finite-range coupling to independent local centers}
\label{sec:tls-finite-range}

We now specialize the general formulas of
Secs.~\ref{sec:deltaJ-formal}–\ref{sec:spectrum} to an environment (compressible strip)
composed of \emph{independent} local degrees of freedom (e.g., TLSs; see Fig.~\ref{fig:TLmodels}(c)) whose
microscopic response is \emph{local} in space, while the edge-environment
coupling is of finite range. We do not need a detailed Hamiltonian of the environment:
it suffices that each site has a local retarded susceptibility
$\chi_{\rm loc}^R(\omega)$ with a gapped low-frequency window,
and that different sites are uncorrelated.
The set-up, disorder averaging, and  the form factor are discussed in Sec.~\ref{sec:deltaJ-formal}, and the following analysis is based on the Eqs.~\eqref{eq:local} and \eqref{eq:S}. We need to stress merely that 
the dependence of local centers implies
\begin{equation}
\chi_q^0\equiv-\,g^2n_s\,|\tilde f(0)|^2\chi_{\rm loc}^R(0)>0,
\label{eq:tls-locality}
\end{equation}
where $n_s$ is the site density and $\chi_q^0$ is the static,
uniform compressibility of the strip (see Sec.~\ref{sec:diffusion}).

\subsection{Small–frequency expansion and the sign of $\Delta J_{\rm th}$}

For $|\omega|\ll v/x_0$, we expand
\begin{align}
{\cal F}(\omega)&=M_1+i\frac{\omega}{v}M_2-\frac{\omega^2}{2v^2}M_3+\cdots,\nonumber
\\[0.3em]
M_n&\equiv\int_{0}^{\infty}\!r^n\,C(r)\,dr>0,\qquad n=0,1,\ldots.
\label{eq:F-expand}
\end{align}
Hence, using Eq.~\eqref{eq:S-finiterange}
\begin{equation}
\mathcal S(\omega)=\frac{n_s g^2}{v^2}
\left[M_1\,\mathrm{Im}\,\chi_{\rm loc}^R(\omega)
+\frac{\omega}{v}\,M_2\,\mathrm{Re}\,\chi_{\rm loc}^R(\omega)\right],
\label{eq:S-expand}
\end{equation}
neglecting terms $\mathcal O(\omega^2)$.
In equilibrium, $\mathrm{Im}\,\chi_{\rm loc}^R(\omega)<0$ for
$\omega>0$ (positive spectral weight). 

If the local response
is gapped, $\mathrm{Im}\,\chi_{\rm loc}^R$ is exponentially
small in the thermal window and the dispersive term dominates.
Expanding the form factor  
$C(k)\equiv|\,\tilde f(k)\,|^2$ near $k=0$ and using 
\eqref{eq:F-expand}
\begin{equation}
C(k)=|\,\tilde f(k)\,|^2
=2M_0 - M_2\,k^2+\cO(k^4),
\label{eq:Ck-expand}
\end{equation}
and using $\mathrm{Re}\,\chi_{\rm loc}^R(0)$ from
\eqref{eq:tls-locality} (i.e., neglecting ${\cal O}(\omega^2)$ corrections) gives the universal linear form
\begin{equation}
\mathcal S(\omega)
=-\,\frac{M_2}{2M_0}\,\frac{\chi_q^0}{v^3}\,\omega,
\qquad
|\omega|,T\ll  v/x_0.
\label{eq:S-linear}
\end{equation}

With the $T{=}0$ piece subtracted, Eq.~\eqref{eq:DJ-master}
yielding the low-$T$ law
\begin{equation}
\Delta J_{\rm th}(T)
=-\,\frac{\pi^2 M_2}{60M_0}\;\frac{\chi_q^0}{v^3}\;T^4,
\qquad T\ll v/x_0.
\label{eq:DeltaJ-TLS}
\end{equation}
The sign and magnitude are controlled solely by the static
compressibility $\chi_0>0$ and the positive geometric moment
$M_2$.

\subsection{Plasmon spectrum: finite range self–energy}

In the present case the spectrum is given by Eqs.~ \eqref{eq:s3-onshell-local} in  Sec.~\ref{sec:spectrum}.
In the gapped window,
$\mathrm{Im}\,\chi_{\rm loc}^R(vk)\approx 0$ and hence $\Gamma_k\approx 0$.
Using Eqs.~\eqref{eq:tls-locality} and \eqref{eq:Ck-expand}, one obtains
\begin{equation}
\delta\omega_k
=-\,\frac{\chi_q^0}{2\pi}\,
\Big(k-\frac{M_2}{2M_0}\,k^3\Big)\;+\;\cO(k^5).
\label{eq:dwk-FF}
\end{equation}
Therefore the group velocity is reduced (negative linear shift), and
the leading nonlinear correction is \emph{convex} (positive $k^3$ term).
Both effects are set by the same static $\chi_q^0$ and the geometry of the
coupling via the positive moments $M_0$ and $M_2$.

It is worth noting that the convex curvature of the plasmon dispersion
[Eq.~\eqref{eq:dwk-FF}], which in the present framework arises from the
dielectric back action of the compressible strip, reproduces the trend
previously identified in Ref.~\cite{IvanRelaxation}.  
In that earlier work, a convex spectrum was shown to correlate with a
reduction of the energy flux carried by the edge plasmon, a qualitative
connection that reappears here despite the different microscopic origin of
the effect.  
To quantify this connection, we introduce the dimensionless ratio
\begin{equation}
\Lambda\equiv-\;\left.\frac{\Delta J_{\rm th}(T)}{\;\omega\,\delta\omega_{\rm nl}(k)\;}\right|_{\,T,\,vk\to\omega},
\label{eq:Lambda}
\end{equation}
where $\delta\omega_{\rm nl}(k)$ denotes the nonlinear part of the correction to the
plasmon dispersion. 
Comparing the $T^{4}$ coefficient in Eq.~\eqref{eq:DeltaJ-TLS} with the
$k^{3}$ coefficient in Eq.~\eqref{eq:dwk-FF} yields a parameter-free universal
value,
\begin{equation}
\Lambda=\Lambda_{\rm FR}
\equiv \frac{\pi^{3}}{15},
\label{eq:universal-relation}
\end{equation}
which provides an experimental cross-check for the scenario of local,
gapped disorder with {\em finite-range} edge coupling.

It is illuminating to contrast this with the case of lateral transport
discussed in the previous section.  
Comparing the $T^{4}$ correction to the heat flux in
Eq.~\eqref{eq:DJ-lowT-q} with the nonlinear spectral correction in
Eq.~\eqref{eq:s4-regC}, one finds
\begin{equation}\Lambda=
\Lambda_{\rm D}
\equiv \frac{\pi^{3}}{30}\sqrt{\frac{\gamma_s}{\omega_*}}
=\frac{\pi^{3}}{30}\frac{\gamma_s l_s}{v},
\label{eq:D-relation}
\end{equation}
where $l_s=\sqrt{D/\gamma_s}$ is the relaxation length.  
In contrast to $\Lambda_{\rm FR}$, this ratio is not universal:
it depends on the dimensionless parameter $(l_s/v)/\tau_s$, i.e., the ratio
of the plasmon flight time $l_s/v$ to the relaxation time $1/\gamma_s$.
This sensitivity to relaxation strength may therefore help distinguish
between local dielectric and diffusive mechanisms in experiment.

\section{Microscopic dipolar model of the dielectric response}
\label{Microscopic model}

In Sec.~\ref{sec:tls-finite-range} we modeled the compressible strip as a set of
independent local centers with site-diagonal susceptibility
$\chi_{ij}^R(\omega)=\delta_{ij}\chi_{\rm loc}^R(\omega)$ and finite-range
edge coupling $g_i(x-x_i)=g f(x-x_i)$.
We now consider a concrete realization of this scenario by placing localized
centers, TLSs in this case, at positions $(x_i,y=d)$ along the edge,
with linear density $n_s=1/a$ and dipole moments
$\boldsymbol{\mu}_i=\mu_i\,\sigma_i^z\,\hat{y}$ oriented perpendicular to the edge
(for simplicity of the following analysis).
Each TLS is characterized by its coupling to the electric field and by its linear response:
\begin{equation}
H_{\rm TLS}=\sum_i (\Delta_i\sigma_i^x-\mu_i\,\sigma_i^z\,E_y),\qquad
\langle \sigma_i^z\rangle=-\,\mu_i\chi_{\rm loc}^R E_y,
\label{eq:RPA-TLS}
\end{equation}
where the response function $\chi_{\rm loc}^R(\omega)=-1/\Delta_i$ is gapped, with
a characteristic energy $\Delta\sim\Delta_i$.
In what follows we first discuss the case of a ``narrow'' compressible strip
of width $W\ll v/\omega$ and then extend the analysis to a wide strip.

\subsection{Narrow compressible strip: $Wk\ll 1$.}

The coupling of TLSs to the edge density is of electrostatic origin
and can be obtained by evaluating the $E_y$ component of the electric field at the
positions of the TLSs:
\begin{equation}
H_{\rm int}= -\frac{d}{4\pi\epsilon}\int dx\sum_i\frac{\mu_i\sigma_i^z\rho(x)}{\big[(x-x_i)^2+d^2\big]^{3/2}},
\end{equation}
where $\rho(x)=\partial_x \phi(x)/2\pi$ is the charge density of the edge plasmon, and
$\epsilon$ is the dielectric constant.
Comparing this to the general coupling $\int dx\, g_i(x-x_i)\,\rho(x)\,S_i$,
we identify
\begin{equation}
g_i(x-x_i)=-\frac{\mu_i}{4\pi\epsilon}\frac{d}{\big[(x-x_i)^2+d^2\big]^{3/2}}.
\end{equation}
Thus, the disorder-averaged coupling constant $g$ and the profile $f(x)$ are
\begin{equation}
g=-\frac{\mu}{4\pi\epsilon},\qquad
f(x)=U(x,d)\equiv\frac{d}{(x^2+d^2)^{3/2}},
\end{equation}
where $\mu$ is the disorder-averaged dipole moment of TLSs.

In Fourier space we obtain
\begin{equation}
\tilde f(k)=U(k,d)=2|k|\,K_{1}(|k|d),
\label{eq:fk}
\end{equation}
where $K_{1}(z)$ is the modified Bessel function of the second kind.
For small argument $z$ one has the asymptotic behavior
\begin{equation}
K_1(z) \simeq \frac{1}{z}
+ \frac{z}{2}\Big[\ln\!\Big(\frac{z}{2}\Big)+\gamma - \frac{1}{2}\Big]
+ \mathcal O\!\big(z^3\ln z\big),
\label{eq:Kz}
\end{equation}
where $\gamma\approx 0.57721$ is the Euler's constant. Therefore, the spectrum in Eqs.~\eqref{eq:s3-onshell-local} acquires a logarithmic
singularity at small $k$ through this asymptotic form.
Physically, it is natural to cut off this logarithm at small argument,
which corresponds to cutting off the original dipolar interaction at long distances,
for instance at the disorder correlation length larger than $d$ or at the distance to a gate.
With such a cutoff the on-shell plasmon shift becomes
\begin{equation}
\delta\omega_k=-\frac{\mu^2n_s}{8\pi^3\epsilon^2d^2\Delta}\,\left(k-c\,d^2k^3\right)+{\cal O}(k^4),
\label{eq:v-correction-narrow}
\end{equation}
where the numerical constant $c>0$ is of order unity, and the spectrum is manifestly convex.

According to Eqs.~\eqref{eq:S}, the kernel $\mathcal{S}(\omega)$ in this dipolar model is
\begin{equation}
\label{eq:S-dip}
\mathcal{S}(\omega)=\frac{\mu^2n_s}{(4\pi\epsilon v)^2\Delta}\,
U(k,d)\,\partial_kU(k,d)\big|_{k\to \omega/v}.
\end{equation}
Using the small-$k$ expansion of $U(k,d)$, the leading low-frequency behavior can be written as
\begin{equation}
\mathcal{S}(\omega)
= -\frac{c\mu^2n_s}{4(\pi\epsilon)^2v^3\Delta}\,\omega
+{\cal O}(\omega^3),\qquad Wk\ll 1.
\label{eq:S-narrow}
\end{equation}
Thus,  the correction to the heat flux is negative and scales as $T^4$,
in agreement with the general result \eqref{eq:S-linear} for a gapped, local environment, and with the same universal ration $\Lambda=\Lambda_{\rm FR}$.

\subsection{Wide compressible strip: $Wk\gg 1$.}

It often happens in experiment that the compressible strip is relatively wide
(with $W$ up to $1\,\mu{\rm m}$), so that $Wk\gg 1$ for the relevant energies.
The results above may then be straightforwardly generalized.
We model the strip as a stack of quasi-1D dipolar lines, separated by a distance $a$ in the transverse
direction, each line having the same linear density $n_s=1/a$ and responding independently.
This amounts to replacing the distance $d\to y$ and integrating
Eqs.~\eqref{eq:s3-onshell-local} and \eqref{eq:S-dip} over $y$ with density $n_s$,
from $y=d$ to $y=W$. In the limit $Wk\gg 1$ the large-argument asymptotic of the Bessel function
\begin{equation}
K_1(z)\simeq \sqrt{\frac{\pi}{2z}}\, e^{-z}\left(1+{\cal O}(z^{-1})\right), \qquad z\gg 1,
\end{equation}
permits to extend the integral to infinity.
For the spectrum we obtain
\begin{equation}
\label{eq:onshell-local-dipolar}
\delta\omega_k=-\frac{\mu^2n_s^2}{8\pi^3\epsilon^2d\Delta}\,\left(k-c'dk^2\right)+{\cal O}(k^3),
\end{equation}
where the constant $c'$ 
\begin{equation}
c'=\int_0^\infty dz\left[\frac{1}{z^2}-K_1^2(z)\right]>0.
\label{eq:c-prime}
\end{equation}
is again of order unity.
Thus, the negative correction to the plasmon group velocity is enhanced by a factor of order
$d/a\gg 1$ compared to the narrow-strip case [Eq.~\eqref{eq:v-correction-narrow}],
and the spectrum remains convex.

For the heat-flux correction in the wide-strip regime, using Eq.~\eqref{eq:S-dip} in the same way,
we arrive at
\begin{equation}
\mathcal{S}(\omega)=-\frac{c'\mu^2n^2_s}{8(\pi\epsilon v)^2\Delta}\sgn(\omega)
+{\cal O}(\omega).
\label{eq:S-wide}
\end{equation}
Using Eq.~\eqref{eq:DJ-master}, the corresponding correction to the energy flux reads
\begin{equation}
\Delta J_{\rm th}
=-\frac{\zeta (3)c'\mu^2n^2_s}{8\pi^4(\epsilon v)^2\Delta}\, T^3
+{\cal O}(T^4),\qquad Wk\gg 1.
\label{eq:th-wide}
\end{equation}
Finally, it is instructive to compare the correction to the heat flux with the nonlinear
correction to the plasmon spectrum:
\begin{equation}
\Lambda=\Lambda^{(1)}_{\rm DIP}\equiv \frac{\zeta (3)}{\pi }.
\label{eq:universal-relation2}
\end{equation}
We again find a universal ratio, which has to be compared to the result \eqref{eq:universal-relation}, for the finite-range coupling.

The above analysis also shows that the exponent $\alpha$ in the power-law scaling, $-\Delta J_{\rm th}\propto T^\alpha$, can be further reduced once one accounts for the fact that the disorder density in the compressible strip varies smoothly across its width.
Indeed, if the local density of dipolar centers increases gradually from the edge and reaches its maximum value $n_s^2$ near the middle of the strip, then the transverse integration acquires an additional weight at larger y. For example, growing linearly from $y=d$ to $y=d+W/2$ this enhanced contribution modifies the low-frequency kernel $\mathcal S(\omega)$ in the regime $d \ll k^{-1} \ll W$,
\begin{equation}
\mathcal{S}(\omega)=-\frac{\mu^2n^2_s}{4(\pi\epsilon)^2Wv\Delta}\,\frac{1}{\omega}+{\cal O}(\omega^0),
\label{eq:S-wide-2}
\end{equation}
such that
\begin{equation}
\Delta J_{\rm th}
=-\frac{\mu^2n^2_s}{48(\pi\epsilon)^2Wv\Delta}\, T^2
+{\cal O}(T^3),\qquad Wk\gg 1,
\label{eq:th-wide-2}
\end{equation}
i.e., a power-law exponent reduced from 3 to 2.
This illustrates how the broad spatial profile of disorder within a wide compressible strip can soften the temperature dependence of the missing-heat anomaly even further.

\section{Discussion}
\label{sec:discussion}

We now discuss the physical meaning of the effect. 
In essence, we propose two complementary scenarios that may account for the suppression of the heat flux below its quantum value, both arising from the polarization-type (dielectric or hydrodynamic) response of the compressible strip at the QH edge (as sketched in Fig.~\ref{fig:TLmodels}(c)). 
Despite their different phenomenology, both mechanisms rely on the same underlying principle: a \emph{negative back action} exerted by disorder-induced polarization in the compressible strip onto the propagating edge mode.

In the fast (local, gapped) regime, the physical picture can be summarized as follows:  
An edge plasmon produces a local fluctuation of the charge density. 
This excess charge, in turn, polarizes nearby localized degrees of freedom within the compressible strip. 
Because these degrees of freedom couple nonlocally to the edge, their activation generates an attractive response downstream from the excitation point, effectively pulling the excess charge away and thereby reducing the local charge fluctuation at the edge. 
This negative feedback constitutes the origin of the reduced heat transport.

A more microscopic realization of the dielectric scenario arises in the
\emph{dipolar model} developed in Sec.~\ref{Microscopic model}, where the edge
plasmon couples electrostatically to localized dipole moments embedded in the
compressible strip.  
Such long-range dipolar interactions substantially enhance the nonlocal
dielectric response, particularly when the strip is wide.  
Consequently, the back action modifies both the plasmon spectrum and the heat
flux more strongly than in the generic finite-range model, yielding not only
the universal $T^{4}$ suppression for a narrow strip ($Wk\!\ll\!1$), but also a
 $T^{3}$ law in the wide-strip regime ($Wk\!\gg\!1$).  
If the disorder density varies smoothly across the strip, the transverse
integration becomes even more infrared-sensitive, and the resulting correction
may soften further to a $T^{\alpha}$ dependence with $\alpha=2$ or even lower.

By contrast, in the slow diffusive regime,
the continuous version of this scenario emerges when the polarization field \(q(x,t)\) within the strip can propagate laterally. 
If this transport is cut off by relaxation at a rate \(\gamma_s\), the resulting feedback remains negative but acquires an extended spatial range, leading to enhanced nonlocality in the edge response.
If the lateral transport is dominated by a diffusion pole, the back action changes sign. 
In that case, the correction to the heat flux becomes positive. 
Such behavior naturally occurs for conserved quantities, for example, for the charge response of the compressible strip. 
In this regime, the temperature dependence of the correction agrees with that found in Ref.~\cite{stabler_nonlocal}, where a similar mechanism was analyzed using the transmission line model.

To assess the feasibility of the proposed mechanisms, we compare  
the magnitude of the correction \eqref{eq:DeltaJ-TLS} to the quantum of the heat flux  \eqref{eq:Jq} yields
\begin{equation}
\frac{|\Delta J_{\rm th}(T)|}{J_q(T)}\sim\frac{|\Delta v|}{v}\,\frac{x_0^2\,T^2}{v^2}\,,
\end{equation}
where \(\Delta v=-\chi_q^0/2\pi\) denotes the plasmon-velocity correction from Eq.~\eqref{eq:dwk-FF}. 
Taking \(x_0\) to be of the order of the compressible-strip width (\(x_0\!\sim\!10^{-6}\,\mathrm{m}\)), \(v = 10^5\,\mathrm{m/s}\) for the plasmon velocity, and an effective temperature \(T = 50~\mu\mathrm{eV}\) (as in Ref.~\cite{Sueur}), one obtains \(x_0^2\,T^2/v^2 \sim 1\). 
Thus, if the plasmon-velocity renormalization is appreciable, \( |\Delta v|/v \sim 1 \), the effect could indeed account for the experimentally observed suppression of the energy flux.

Further insight into the strength of the effect can be obtained from the
microscopic dipolar model developed in Sec.~\ref{Microscopic model}.  
Using Eq.~\eqref{eq:th-wide}, the relative correction to the heat flux can be
expressed as
\begin{equation}
\frac{|\Delta J_{\rm th}(T)|}{J_q(T)}
=\frac{3\zeta (3)\,\mu^2 n_s^2}{2\pi^5 (\epsilon v)^2 \Delta}\,T
\;\sim\;\mu^2 n_s^2\,T/\Delta,
\end{equation}
where in the last step we substituted the edge–plasmon velocity 
$v = 1/(4\pi^2\epsilon)$~\cite{VolkovMikhailov1988}.  
For smooth disorder it is natural to expect $\mu n_s = \mu/a \sim 1$.  
While our analysis assumes coupling to gapped local centers, and hence 
$T \ll \Delta$, using $T/\Delta \sim 1$ provides an estimate of the maximal
possible suppression.  
Even under conservative assumptions, the resulting magnitude of the effect is
sizable, indicating that the dipolar-disorder mechanism is sufficiently strong
to account for the experimentally observed suppression of the heat flux.

Determining whether the lateral transport of the polarization field is active in experiment requires further investigation, as it depends sensitively on the disorder density and the level spacing of localized states in the strip. 
Assuming, however, that such transport occurs, it should enhance the effect. 
Denoting by \(\Delta J_{\rm th}^{\rm nl}\) the nonlocal correction [Eq.~\eqref{eq:DJ-lowT-q}] and by \(\Delta J_{\rm th}^{\rm loc}\) the local one [Eq.~\eqref{eq:DeltaJ-TLS}], we find
\begin{equation}
\frac{\Delta J^{\rm nl}_{\rm th}}{\Delta J^{\rm loc}_{\rm th}}=\frac{M_0}{M_2}\,
\frac{v\sqrt{D}}{\gamma_s^{3/2}}\propto 
\frac{v\,l_s}{x_0^2\gamma_s}\,,
\label{eq:lateral-transport}
\end{equation}
where \(l_s = \sqrt{D/\gamma_s}\) is the relaxation length [see Eq.~\eqref{eq:chiq-x}]. 
Using the minimal relaxation rate \(\gamma_s \sim T\) from Eq.~\eqref{eq:DJ-lowT-q} and the experimental parameters of Ref.~\cite{Sueur}, we estimate \(v/\gamma_s \sim 10^{-6}\,\mathrm{m}\). 
Assuming \(x_0 \sim 10^{-6}\,\mathrm{m}\) and an efficient lateral transport regime with \(l_s \gg x_0\), we obtain $\Delta J^{\rm nl}_{\rm th}\gg\Delta J^{\rm loc}_{\rm th}$ indicating a substantial enhancement of the missing-heat effect due to the nonlocal propagation of polarization within the compressible strip.

An important diagnostic of the underlying mechanism is provided by the
\emph{universality relations} connecting the correction to the heat flux and
the nonlinear part of the plasmon spectrum.  
Each regime identified in this work is characterized by a distinct,
dimensionless ratio \eqref{eq:Lambda}.  
For the local finite-range local model this ratio is the universal number
$\pi^{3}/15$ [Eq.~\eqref{eq:universal-relation}], whereas in the diffusive
(hydrodynamic) regime the corresponding relation depends parametrically on the
ratio $\sqrt{\gamma_s/\omega_*}$ [Eq.~\eqref{eq:D-relation}].  
In the dipolar model, a wide compressible strip produces yet another universal
value, $\zeta(3)/\pi$ [Eq.~\eqref{eq:universal-relation2}], reflecting the
strong nonlocality of the dipolar coupling.  
Since these ratios link two independent observables, heat transport and the
curvature of the plasmon dispersion, they provide a powerful experimental tool
for identifying which physical mechanism dominates in a given device or
temperature range.

\section{Conclusion}
\label{sec:conclusion}

We have developed a unified perturbation theory framework describing a chiral quantum Hall edge mode coupled to a disordered, compressible strip acting as an ``environment''.  
Within this formulation, all interaction effects are contained in the retarded susceptibility of the environmental density field $q(x,t)$, which governs both (i) the correction to the heat flux through the upstream-downstream back action kernel $\mathcal S(\omega)$ [Eq.~\eqref{eq:S-kernel}] and (ii) the plasmon self-energy $\Sigma^{R}(k,\omega)$ [Eq.~\eqref{eq:s3-GR-sigma}].  
This establishes a single theoretical language connecting the modification of energy transport with the renormalization of the edge excitation spectrum.

For a model of \emph{local} microscopic centers with a \emph{finite-range}
coupling to the edge plasmon, the low-frequency kernel is linear and odd in
$\omega$ [Eq.~\eqref{eq:S-linear}], leading to a negative $T^4$
correction to the quantum heat flux [Eq.~\eqref{eq:DeltaJ-TLS}].  
In the same regime, the on-shell plasmon spectrum exhibits a negative velocity
renormalization together with a convex $k^{3}$ correction
[Eq.~\eqref{eq:dwk-FF}].  
These two manifestations of dielectric back-action are not independent: they
are tied by a parameter-free universality relation
[Eq.~\eqref{eq:universal-relation}], providing a direct and experimentally
accessible link between nonlinear spectral curvature and deviations from
quantized heat transport.

A more microscopic realization of the dielectric scenario is provided by the
dipolar model of Sec.~\ref{Microscopic model}, where the edge plasmon couples
electrostatically to localized dipoles inside the compressible strip.  
This long-range interaction enhances nonlocality and modifies both the spectrum
and the heat flux more strongly than the generic finite-range model.  
For a narrow strip ($Wk\!\ll\!1$) it yields the same $T^{4}$ law,
whereas for a wide strip ($Wk\!\gg\!1$) the transverse integration produces a
distinct $T^{3}$ behavior, which can soften to $T^{2}$ if the disorder density
varies smoothly across the strip.  
The wide strip dipolar regime also exhibits its own universal
transport-spectrum ratio \eqref{eq:universal-relation2}, complementing the
finite-range value \eqref{eq:universal-relation} and the hydrodynamic one \eqref{eq:D-relation}.  
Together, these characteristic ratios provide a practical way to identify the
dominant mechanism behind the observed heat flux suppression.

In the regime where the compressible strip supports diffusive transport, the back action is fully determined by the hydrodynamic response function $\chi_q^R(k,\omega)$ [Eq.~\eqref{eq:chiq-hydro}].  
At the lowest temperatures, relaxation with a finite rate $\gamma_s>0$ yields a negative $T^4$ correction to the heat flux [Eq.~\eqref{eq:DJ-lowT-q}].
The associated plasmon dispersion and damping follow from Eqs.~\eqref{eq:s4-dw}-\eqref{eq:s4-regC}, exhibiting a negative velocity renormalization and a convex $k^3$ correction \eqref{eq:s4-regC}. In the limit of conserved dynamics ($\gamma_s\to0$) the correction changes sign and crosses over to the diffusion-controlled power laws $T^{5/2}$ and $T^{3/2}$ [Eqs.~\eqref{eq:DJ-midT-q}–\eqref{eq:DJ-highT-q}].

Using parameters characteristic of experiments on the heat and charge transport along quantum Hall edges (for example, Ref.~\cite{Sueur}), the relative correction to the heat flux $|\Delta J_{\rm th}|/J_q$ can naturally be of order unity.  
This occurs when the interaction range is comparable to the width of the compressible strip ($x_0\sim1~\mu\mathrm{m}$), the plasmon velocity is $v\sim10^5$-$10^6~\mathrm{m/s}$, and the effective temperature (or bias) is $T\sim50~\mu\mathrm{eV}$.  
The resulting sign of the correction agrees with the experimental observation.  
Moreover, lateral transport of polarization may considerably amplify the effect, with the enhancement factor scaling as $v\,l_s/(x_0^2\gamma_s)$, where $l_s=\sqrt{D/\gamma_s}$ is the relaxation length [Eq.~\eqref{eq:lateral-transport}].

In summary, the response of the compressible strip, described generically by its retarded susceptibility $\chi_q^R(k,\omega)$ and encompassing both local dielectric and diffusive limits, provides a coherent and quantitatively plausible explanation of the observed partial suppression of the heat flux quantization at QH edges reported in Ref.~\cite{Sueur}.
Beyond resolving this ``missing heat flux’’ anomaly, the framework unifies the thermal and spectral manifestations of dissipation, enabling systematic exploration of microscopic energy transport in chiral quantum channels.

\section*{Acknowledgments}     
We acknowledge the financial support from the Swiss National Science Foundation. 

\bibliographystyle{apsrev4-2}
\bibliography{References,refs-2,refs,Paper}

\appendix

\section{Free correlators and the fluctuation–dissipation theorem}
\label{app:FDT}

This appendix provides the basic correlation functions for the free chiral edge.
We derive the retarded ($R$), advanced ($A$), and Keldysh ($K$) correlators explicitly,
establish their symmetry relations, and show how they are connected by the
fluctuation-dissipation theorem (FDT).  
All conventions here are those used in
Secs.~\ref{sec:model}–\ref{sec:deltaJ-formal}.


We consider a single right-moving chiral boson $\phi(x,t)$ with velocity $v>0$ and
Hamiltonian
\begin{subequations}
\begin{align}
H_{\rm edge}&=\frac{v}{4\pi}\int dx\,\big(\partial_x\phi(x)\big)^2,
\label{app:eq:Hedge}
\\[0.4em]
[\partial_x\phi(x),\phi(y)]&=2\pi i\,\delta(x-y).
\label{app:eq:comm}
\end{align}
\end{subequations}
The equation of motion $(\partial_t+v\partial_x)\phi=0$ confirms
that excitations propagate to increasing~$x$.

It is convenient to use the standard mode expansion for a right-moving field:
\begin{subequations}
\label{app:eq:mode}
\begin{align}
\phi(x,t)=
\int_0^{\infty}\frac{dk}{\sqrt{k}}\,
\Big[b_k\,e^{ik(x-vt)}+b_k^\dagger\,e^{-ik(x-vt)}\Big],
\label{app:eq:mode-a}
\\[0.4em]
[b_k,b_{k'}^\dagger]=\delta(k-k').
\label{app:eq:mode-b}
\end{align}
\end{subequations}
Thermal expectation values at temperature $T_e=1/\beta_e$ are
\begin{equation}
\langle b_k^\dagger b_{k'}\rangle
=\delta(k-k')\,n_B(vk),\qquad
n_B(\omega)=\frac{1}{e^{\beta_e\omega}-1}.
\label{app:eq:nB}
\end{equation}


The retarded and advanced correlators are defined as
\begin{subequations}
\label{app:eq:GRGA-def}
\begin{align}
G_0^R(x,t)&=-i\,\theta(t)\,\langle[\phi(x,t),\phi(0,0)]\rangle,
\label{app:eq:GRGA-def-a}
\\[0.4em]
G_0^A(x,t)&=+i\,\theta(-t)\,\langle[\phi(x,t),\phi(0,0)]\rangle.
\label{app:eq:GRGA-def-b}
\end{align}
\end{subequations}
Using \eqref{app:eq:mode} and the canonical commutator for $b_k,b_{k'}^\dagger$,
one obtains the equal-time relation \eqref{app:eq:comm}. After time evolution this gives
\begin{equation}
[\phi(x,t),\phi(0,0)]=i\pi\,\mathrm{sgn}(x-vt).
\label{eq:commutator}
\end{equation}
Substituting the result to \eqref{app:eq:GRGA-def} we obtain
\begin{subequations}
\label{app:eq:GRGA-time}
\begin{align}
G_0^R(x,t)&=\pi\,\theta(t)\,\mathrm{sgn}(x-vt),
\label{app:eq:GRGA-time-a}
\\[0.4em]
G_0^A(x,t)&=-\pi\,\theta(-t)\,\mathrm{sgn}(x-vt).
\label{app:eq:GRGA-time-b}
\end{align}
\end{subequations}

Perfoming the Fourier transforming with respect to~$t$ yields
\begin{subequations}
\label{app:eq:GRGA-omega}
\begin{align}
G_0^R(x,\omega)&=\frac{\pi}{i\,\Omega_R}\Big[1+2\theta(x)(e^{i\Omega_R x/v}-1)\Big],
\label{app:eq:GRGA-omega-a}
\\[0.4em]
G_0^A(x,\omega)&=-\frac{\pi}{i\,\Omega_A}\Big[1+2\theta(-x)(e^{i\Omega_A x/v}-1)\Big],
\label{app:eq:GRGA-omega-b}
\end{align}
\end{subequations}
where $\Omega_{R/A}=\omega\pm i0^+$


The Keldysh (or symmetrized) correlator is defined as
\begin{equation}
G_0^K(x,\omega)
=\int dt\,e^{i\omega t}\,\langle\{\phi(x,t),\phi(0,0)\}\rangle.
\label{app:eq:GK-def}
\end{equation}
Using the mode expansion \eqref{app:eq:mode} and the Bose occupation
\eqref{app:eq:nB}, one finds after a short calculation:
\begin{equation}
G_0^K(x,\omega)
=\frac{2\pi}{\omega}\,\coth\!\Big(\frac{\beta_e\omega}{2}\Big)
\,e^{i\omega x/v}.
\label{app:eq:GK-result}
\end{equation}

\end{document}